\documentclass{osa-article}

\journal{oe}
\articletype{Research Article}
\begin{document}

\title{Investigation of the quality of an As35S65 grating by spectroscopic ellipsometry}

\author{Roman Antos,\authormark{1,*} Jan Mistrik,\authormark{2,3} Karel~Palka,\authormark{3,4} Stanislav Slang,\authormark{3} Josef Navratil,\authormark{1} Jaroslav~Hamrle,\authormark{1} Martin Veis,\authormark{1} and Miroslav Vlcek\authormark{3}}

\address{\authormark{1}Institute of Physics, Faculty of Mathematics and Physics, Charles University, Ke Karlovu 5, 12116 Prague~2\\
\authormark{2}Institute of Applied Physics and Mathematics, Faculty of Chemical Technology, University of Pardubice, Studentska 95, Pardubice 532 10, Czech Republic\\
\authormark{3}Center of Materials and Nanotechnologies, Faculty of Chemical Technology, University of Pardubice, Studentska 95, Pardubice 532 10, Czech Republic\\
\authormark{4}Department of General and Inorganic Chemistry, Faculty of Chemical Technology, University of Pardubice, Studentska 95, Pardubice 532 10, Czech Republic
}

\email{\authormark{*}antos@karlov.mff.cuni.cz} %% email address is required



\begin{abstract}
The quality of an $\mathrm{As_{35}S_{65}}$ chalcogenide glass (ChG) grating fabricated by electron beam lithography (EBL) was characterized by optical scatterometry based on spectroscopic ellipsometry (SE) in the visible and near infrared spectral range and complementary techniques providing direct images, especially atomic force microscopy (AFM). The geometric dimensions and the shape of patterned grating lines were determined by fitting modeled values (calculated by the Fourier modal method) to SE experimental data. A simple power-dependent function with only one variable parameter was successfully used to describe the shape of the patterned lines. The result yielded by SE is shown to correspond to AFM measurement with high accuracy, provided that optical constants of ChG modified by EBL were used in the fitting procedure. The line edge roughness (LER) of the grating was also investigated by further fitting the SE data to find out that no LER is optically detectable in the spectral range used, which is essential for the functionality of optical tools fabricated by EBL.
\end{abstract}

%%%%%%%%%%%%%%%%%%%%%%%%%%  body  %%%%%%%%%%%%%%%%%%%%%%%%%%
\section{Introduction}
Chalcogenide glasses (ChGs) are materials suitable for fabrication of various optical elements for light guiding (e.g., waveguides or fibers) or filtering (e.g., diffraction grating spectrometer), especially for infrared (IR) region~\cite{Borisova_1981,Tanaka_2011,Zoubir_OL_2004}. They have a high refractive index and a wide IR transmission window, and are sensitive when exposed to electromagnetic or electron beam radiation. Exposure of ChGs to an electron beam can lead to structural modification, which---together with good solubility in organic or inorganic solutions---makes them usable as high-resolution inorganic resists~\cite{Singh_APL_1982a,Singh_APL_1982b,Teteris_JNCS_2002}. Due to the structural modification, changes of optical and morphological properties can also take place, e.g., a change of refractive index~\cite{Suhara_JJAP_1975,Nordman_JAP_1998,Sarkar_RPhCh_2014}.

Many optical tools made of ChGs require their deposition as a thin film~\cite{Musgraves_AM_2011}, which can then be further modified by lithographic patterning~\cite{Slang_MChPh_2018}. For proper functionality high precision and knowledge of the optical and geometric properties of such patterns are necessary. Optical scatterometry based on spectroscopic ellipsometry (SE)~\cite{Veis_JN_2013,Madsen_STMP_2016}, which is a characterization method used in this paper, is capable of monitoring both optical and geometric properties with high precision, including pattern dimensions (the period, linewidth, or depth~\cite{Antos_JAP_2006}), and also surface oxide overlayers~\cite{Antos_APL_2005} and the precise shape and roughness of patterned line edges~\cite{Antos_OE_2005}. Therefore, SE is very useful for the verification of those pattern properties that are important for the optical functionality of the tools fabricated by lithography.

We here investigate the optical properties and the precise relief of an $\mathrm{As_{35}S_{65}}$ ChG grating fabricated by electron beam lithography (EBL) by means of SE with relatively narrow spectral range of visible and near IR light. We demonstrate that even the SE measurement with the limited range is capable of very accurate determination of the grating profile, and is also very sensitive to the optical properties (refractive index and extinction coefficient in the spectral range used), the substrate's surface oxide, and line edge roughness (LER). The results of the surface profile geometry is also successfully compared with measurements of complementary techniques, atomic force microscopy (AFM), scanning electron microscopy (SEM), and optical microscopy.

\section{Experimental details}
Source bulk $\mathrm{As}_{35}\mathrm{S}_{65}$ ChG was prepared by standard melt-quenching method. High purity 5N elements were weighted into clean quartz ampule in calculated amounts and sealed under vacuum ($\approx\!10^{-3}\ \mathrm{Pa}$). Loaded reagents were melted in a rocking tube furnace at $800\ ^{\circ}\mathrm{C}$ for 48~hours. The ampule with melted glass was quenched in cold water. Thin films were deposited by vacuum thermal evaporation technique (device UP-858, Tesla corp.). The films were prepared from source ChG bulk glass by evaporation from a molybdenum boat onto rotating silicon substrates with evaporation rate $\approx\!1\ \mathrm{nm\,s^{-1}}$ at a residual pressure of $\approx\!10^{-3}\ \mathrm{Pa}$. The film's intended nominal thickness was $\approx\!100\ \mathrm{nm}$. The thickness and evaporation rate were measured using quartz crystal microbalance method (device MSV-1843/A MIKI-FFV).

Electron beam lithography exposure was performed using scanning electron microscope LYRA~3 (Tescan). The latent images of gratings were recorded at 5~kV acceleration voltage and $300\ \mathrm{\mu C/cm}^2$ electron beam dose. The $\mathrm{As}_{35}\mathrm{S}_{65}$ ChG thin films with recorded latent images were etched for 50~s in 0.05~vol.~\% solution of n-butylamine in N,N-dimethylfromamide at $25\ ^{\circ}\mathrm{C}$ in order to remove unexposed thin film material.

\begin{figure}[h!]
\centering\includegraphics[width=11cm]{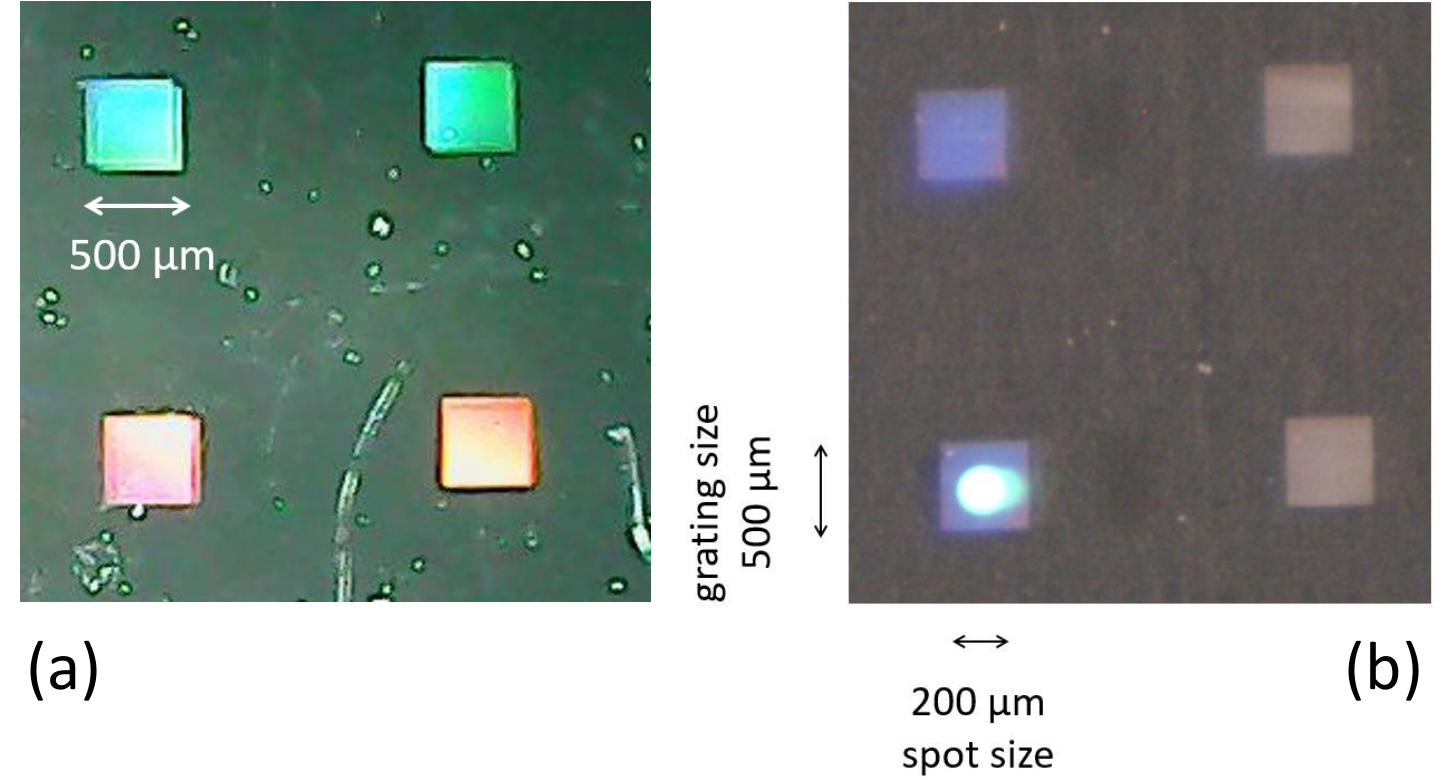}
\caption{Photographs of sample gratings and the focused light spot of the ellipsometer's beam.}
\label{fig:photos}
\end{figure}

\begin{figure}[ht!]
\centering\includegraphics[width=12cm]{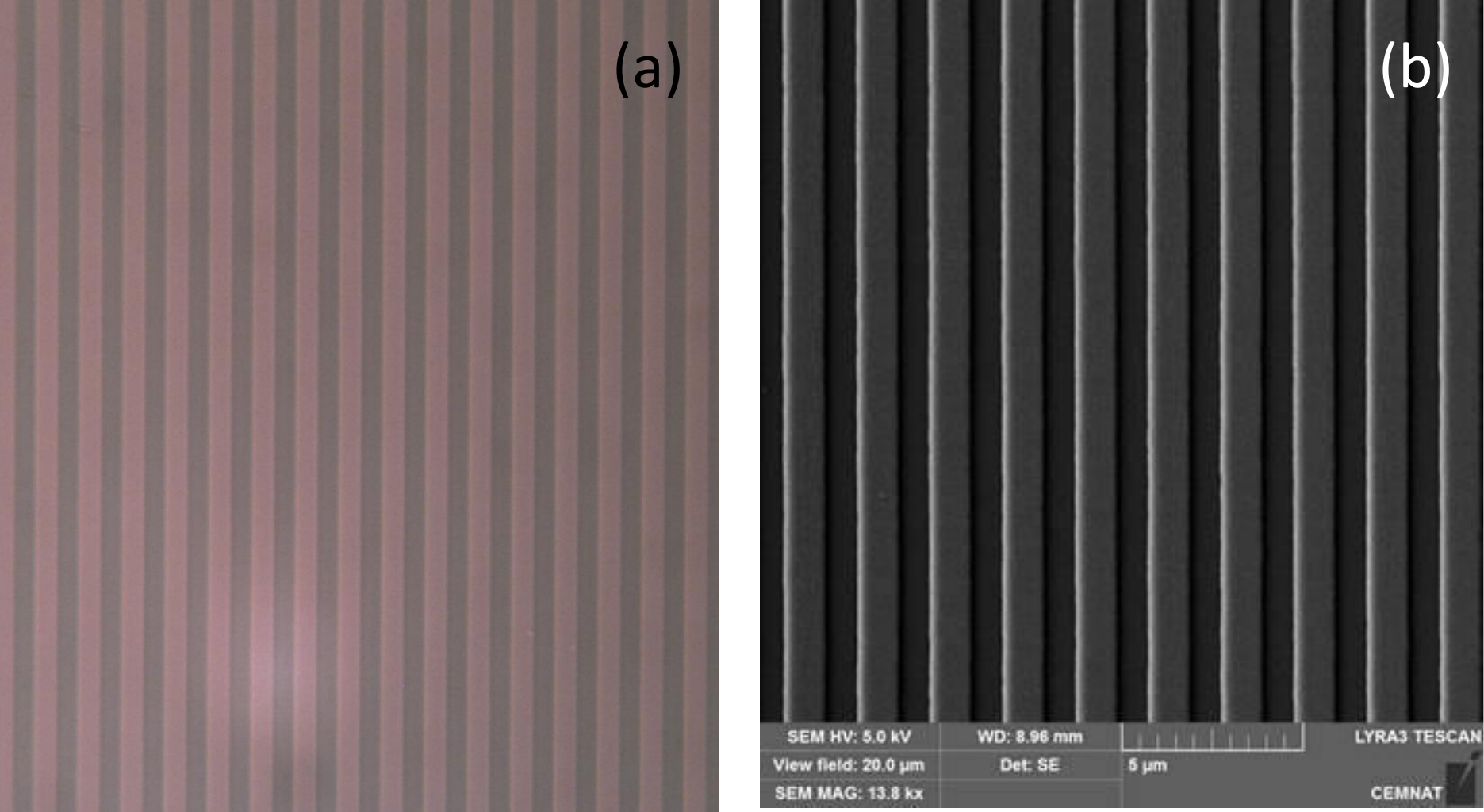}
\caption{Optical micrograph (a) and scanning electron micrograph (b) of the grating.}
\label{fig:micro}
\end{figure}

\begin{figure}[ht!]
\centering\includegraphics[width=9cm]{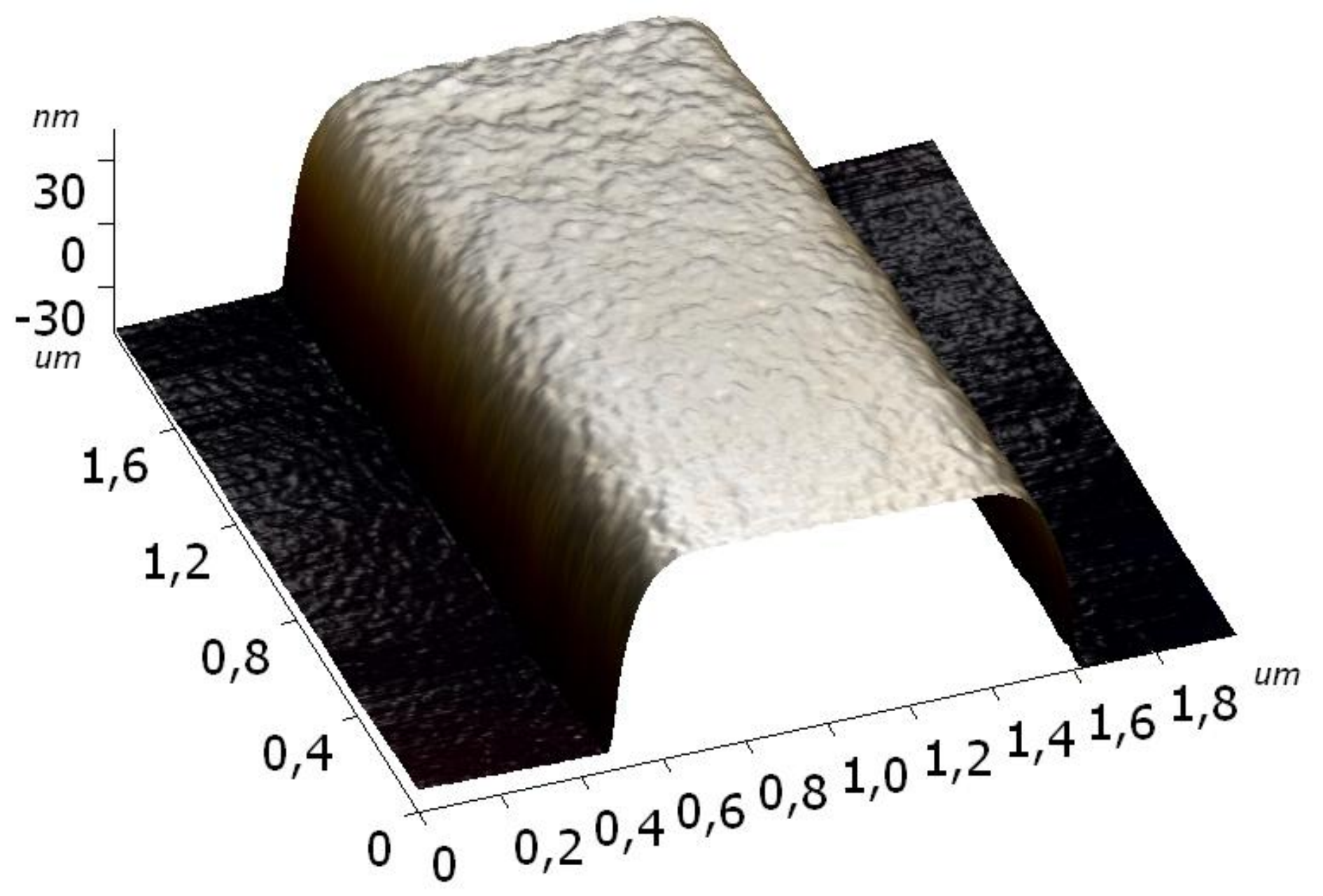}
\caption{Three-dimensional atomic force micrograph of one grating period.}
\label{fig:AFM_3D}
\end{figure}

The fabricated gratings were measured by a spectroscopic ellipsometer (WVASE Woollam, Co.~Ltd.) on a specularly reflected beam. Ellipsometric parameters were recorded with microspot optics due to the limited size of the patterned grating area ($500\times500\ \mathrm{\mu m}^2$, as visible in Fig.~\ref{fig:photos}) in the spectral range from 400 to 1000~nm and for various angles of incidence (20--40$^{\circ}$). The surface profiles were also characterized by complementary measurements of optical microscopy, SEM, and AFM (Figs.~\ref{fig:micro} and~\ref{fig:AFM_3D}). The AFM pictures of various $2\times2\ \mathrm{\mu m}^2$ grating areas were scanned by an NT-MDT NREGRA instrument. SEM images of the sizes $20\times20$, $60\times60$, and $560\times560\ \mathrm{\mu m}^2$ were recorded by LYRA 3 (Tescan). Optical micrographs were recoded by an optical microscope Leitz, Ergolux (mag.~500x).

\section{Optical modeling and data processing}
The optical reflectance coefficients $r_s$ and $r_p$ for the $s$- and $p$-polarized incident waves, respectively, were calculated by the Fourier modal method (FMM) with the implementation described in~\cite{Antos_JAP_2006}. The sample was divided into $N$ slices, one of which was the $\mathrm{SiO}_2$ native layer on the top of the c-Si substrate and $N-1$ of which were ultratnin slices of the ChG grating layer of equal thickness. Each slice was assumed uniform along the vertical axis, so that eigenmodes and their propagation numbers could be calculated from the matrix form of the wave equation. Standard boundary conditions of tangential electric and magnetic field components between adjacent slices were assumed. Proper Fourier factorization of the constitutive relation $\mathbf{D}=\varepsilon\mathbf{E}$ took place as well.

All Fourier and pseudo-Fourier expansions of the fields and permittivity lateral dependence were truncated to have summations $\sum_{n=-n_{\max}}^{n_{\max}}$, where $n_{\max}$ denotes the truncation number. All simulations presented in this paper were carried out with $N=n_{\max}=20$, which was sufficient for the sample under study. All matrices treated by linear-algebraic operations were thus of the order of $2n_{\max}+1=41$.

To investigate the quality of the grating pattern with respect to LER, a quality factor was defined as a reduction of the edge-diffracted light in the amplitude reflectance coefficients, as suggested in~\cite{Antos_OE_2005}. The LER-modified coefficients were defined as
\begin{eqnarray}
    r'_s &=& r_s^{\mathrm{loc}} + \eta(r_s - r_s^{\mathrm{loc}}),\label{eq:LER1}\\
    r'_p &=& r_p^{\mathrm{loc}} + \eta(r_p - r_s^{\mathrm{loc}}),\label{eq:LER2}
\end{eqnarray}
where $r_s$, $r_p$ are reflectance coefficients accurately calculated by FMM, $\eta$ is the quality factor, and
\begin{eqnarray}
    r_s^{\mathrm{loc}} &=& w_{\mathrm{B}}r_s^{(1)} + (1-w_{\mathrm{B}})r_s^{(2)},\\
    r_p^{\mathrm{loc}} &=& w_{\mathrm{B}}r_p^{(1)} + (1-w_{\mathrm{B}})r_p^{(2)}
\end{eqnarray}
are reflectance coefficients in the zeroth-order diffraction calculated by the local modes method, demonstrated in~\cite{Antos_JMMM_2004,Antos_APL_2005}. Here $r_s^{(1)}$, $r_p^{(1)}$ are coefficients of the unpatterned part (i.e., the uniform ChG film), and $r_s^{(2)}$, $r_p^{(2)}$ of the patterned part (i.e., the bare substrate with an air layer replacing ChG). The $r_s^{\mathrm{loc}}$, $r_p^{\mathrm{loc}}$ coefficients are thus the weighted arithmetic mean whose weight is the grating bottom filling factor~$w_{\mathrm{B}}$. According to this definition, the quality factor $\eta$ is a non-negative value between 0~and~1, where $\eta=1$ corresponds to a high-quality grating whose LER cannot be optically detected, while $\eta=0$ corresponds to a grating with so high irregularities that the line edges do not contribute to the diffraction in the zeroth order.

The ellipsometric parameters $\Psi$ and $\Delta$ were calculated from the reflectance coefficients via the definition
\begin{equation}
    \tan\Psi\exp i\Delta = \frac{r_p}{r_s}.
\end{equation}
The obtained modeled values ($\Psi_{\mathrm{m},j}$, $\Delta_{\mathrm{m},j}$) were fitted to the experimental ones ($\Psi_{\mathrm{e},j}$, $\Delta_{\mathrm{e},j}$) by gradually changing grating shape and spatial dimensions to minimize their difference, which was during the fitting evaluated as the angular distance between the experimental and modeled points plotted on Poincar\'{e}'s sphere. This distance is specified by the azimuthal angle $2\Psi$ and the polar angle $\Delta$, i.e.,
\begin{equation}
    \cos\mathcal{E}_j=\mathbf{S}_{\mathrm{e},j}\cdot\mathbf{S}_{\mathrm{m},j},
\end{equation}
where $\mathcal{E}_j$ is the error of the $j$th value and $\mathbf{S}_{\mathrm{e},j}$ and $\mathbf{S}_{\mathrm{m},j}$ are the Stokes vectors of the $j$th experimental and modeled ellipsometric values, respectively, defined by
\begin{eqnarray}
    \mathbf{S}_{\mathrm{e},j}&=&[\sin2\Psi_{\mathrm{e},j}\cos\Delta_{\mathrm{e},j},
    \ \sin2\Psi_{\mathrm{e},j}\sin\Delta_{\mathrm{e},j},
    \ \cos2\Psi_{\mathrm{e},j}],\\
    \mathbf{S}_{\mathrm{m},j}&=&[\sin2\Psi_{\mathrm{m},j}\cos\Delta_{\mathrm{m},j},
    \ \sin2\Psi_{\mathrm{m},j}\sin\Delta_{\mathrm{m},j},
    \ \cos2\Psi_{\mathrm{m},j}].
\end{eqnarray}
In the case of the fitting procedure, the sum of the least squares, $\mathcal{E}_{\mathrm{LS}}^2=\sum_j\mathcal{E}_j^2$, was minimized. In the resulting tables listed below the average error of $M$ compared values, $\mathcal{E}=(1/M)\sum_{j=1}^M\mathcal{E}_j$, is written.

\begin{figure}[ht!]
\centering\includegraphics[width=12cm]{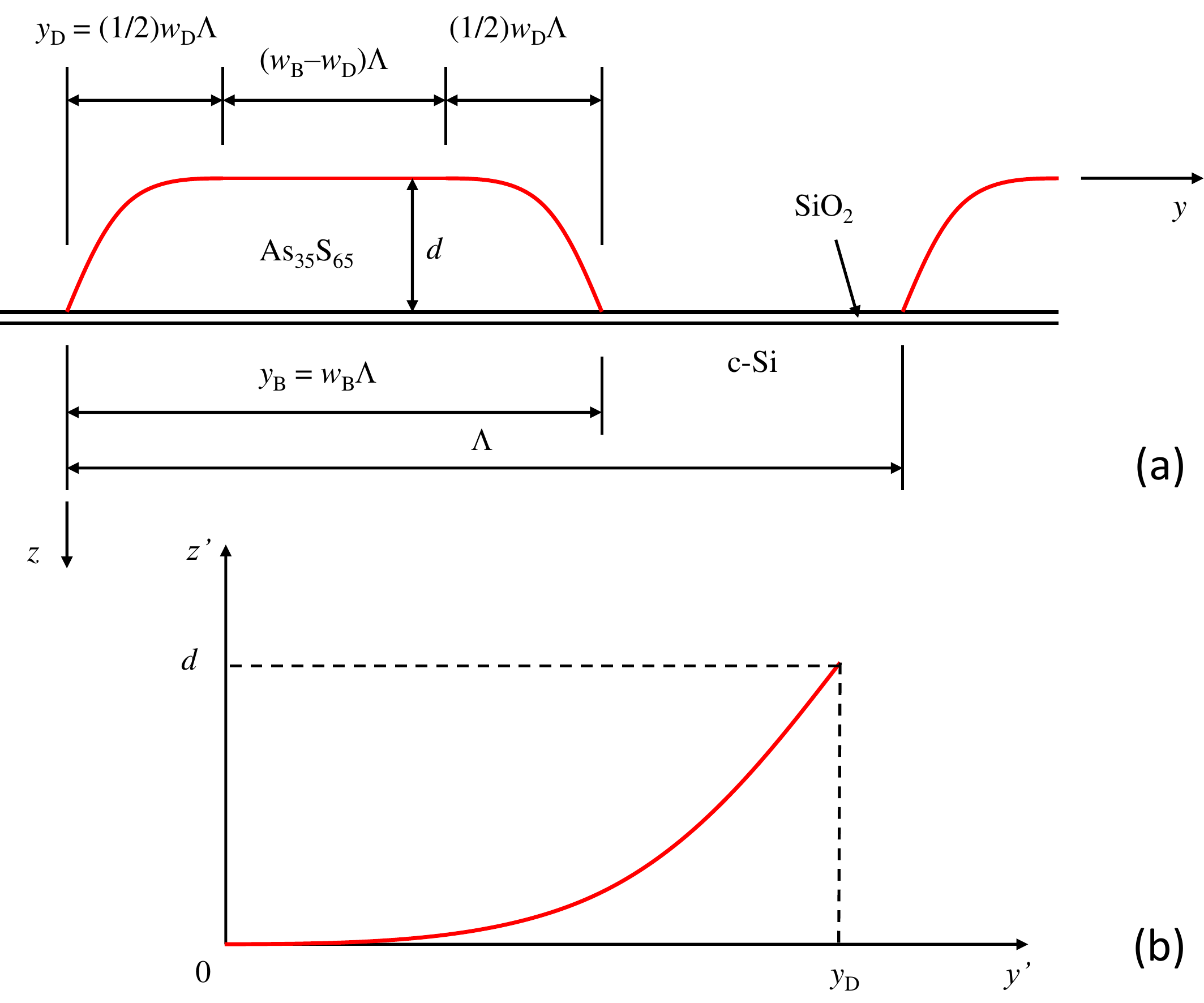}
\caption{Grating geometry and orientation (a) and the function dependence of the patterned edge (b).}
\label{fig:draft}
\end{figure}

The ChG grating and its orientation is displayed in Fig.~\ref{fig:draft}(a). Patterned horizontal lines of ChG, oriented along the $x$-axis and periodic along the $y$-axis, are located on the c-Si substrate, which has a native surface $\mathrm{SiO}_2$ layer developed before the ChG deposition. The period of the grating is denoted~$\Lambda$, the depth~$d$, the bottom linewidth $y_{\mathrm{B}}=w_{\mathrm{B}}\Lambda$, and the top linewidth $y_{\mathrm{T}}=(w_{\mathrm{B}} - w_{\mathrm{D}})\Lambda$. The bottom linewidth therefore overhangs the top one by $y_{\mathrm{D}}=(1/2)w_{\mathrm{D}}\Lambda$ on both sides. The dimensionless values of $w_{\mathrm{B}}$ and $w_{\mathrm{D}}$ denote the corresponding dimensionless filling factors. A detail of the line edge of the width $y_{\mathrm{D}}$ is displayed in Fig.~\ref{fig:draft}(b) with coordinates modified for the purpose of functional parametrization. According the the AFM measurement, a monotonically increasing function is searched for, with the zero derivative at $y'=0$, and with $z'=d$ at $y'=y_{\mathrm{D}}$.

\section{Results and discussion}
\begin{figure}[ht!]
\centering\includegraphics[width=13cm]{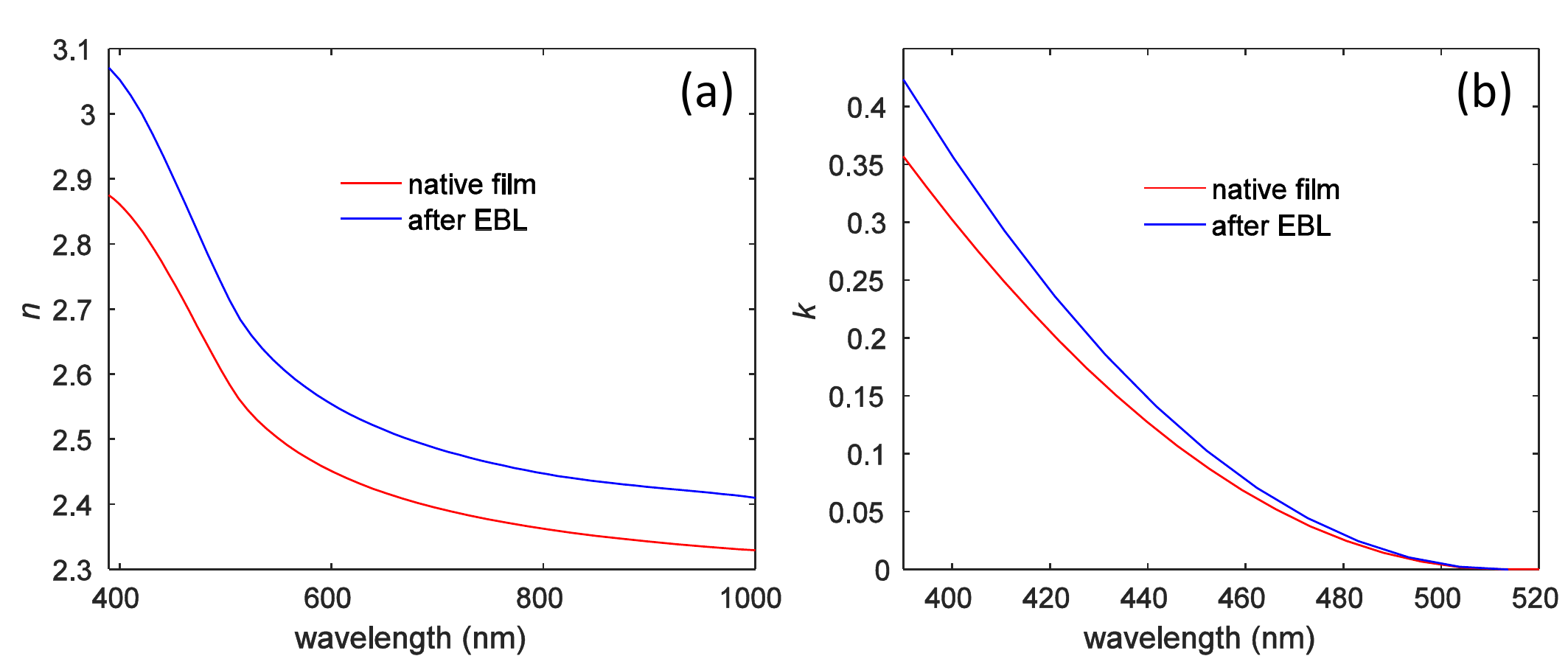}
\caption{Refractive index $n$ (a) and extinction coefficient $k$ (b) of $\mathrm{As}_{35}\mathrm{S}_{65}$ used for the grating pattern. The red curves correspond to native material, the blue curves for material modified by EBL.}\label{fig:AsS_n_k}
\end{figure}

Only geometric dimensions and parameters of the edge shape were determined by fitting the SE data. The optical parameters, refractive index $n$ and extinction coefficient $k$, were determined from SE measurements of an unpatterned thin film reference sample, which was placed in the deposition apparatus beside the sample intended for grating patterning. From the SE fitting procedure carried out on the reference sample, the thickness of the uniform ChG layer $d=106$~nm and the thickness of the native $\mathrm{SiO}_2$ substrate overlayer $d_{\mathrm{SiO}_2}=2$~nm were determined together with the optical parameters of the native ChG displayed in Fig.~\ref{fig:AsS_n_k} (red curves).

For the fitting procedure of the grating, the period $\Lambda=2\,000$~nm, the native $\mathrm{SiO}_2$ substrate overlayer $d_{\mathrm{SiO}_2}=2$~nm, and the optical constants obtained from the native ChG film were fixed. The other geometric parameters, the thickness $d$ of the ChG grating, the filling factor of the bottom linewidth $w_{\mathrm{B}}$, the filling factor of the edges $w_{\mathrm{D}}$, and the edge shape parameters were fitted. Data obtained for five angles of incidence, 20, 25, 30, 35, and~$40^{\circ}$, were fitted together.

The first fitting experiments were carried out with a polynomial dependence (the prime symbols on the modified coordinates are for simplicity omitted)
\begin{equation}\label{eq:polynom1}
    z(y) = a_2y^2 + a_3y^3 + \ldots + a_ny^n,
\end{equation}
where $a_j$ are polynomial coefficients and $n$ is the order of the polynomial. For the fitting procedure, a set of dimensionless parameters $c_j$ was used, so that Eq.~\ref{eq:polynom1} could be rewritten into
\begin{equation}
    \frac{z(y)}{d} = \left(c_2\frac{y}{y_{\mathrm{D}}}\right)^2 + \left(c_3\frac{y}{y_{\mathrm{D}}}\right)^3 + \ldots + \left(c_n\frac{y}{y_{\mathrm{D}}}\right)^n.
\end{equation}
Since the edge curve has to pass through the point $[y_{\mathrm{D}},\ d]$, the parameters are mutually coupled by the formula
\begin{equation}\label{eq:coef_coupling}
1 = c_2^2 + c_3^3 + \ldots + c_n^n,
\end{equation}
so that only $n-2$ parameters are fitted. The results of the fitting for selected orders of the polynomials, $n\in\{2,\ 3,\ 6,\ 8\}$, are shown in Table~\ref{tab:polynom}. The polynomial coefficients written in parentheses were not fitted; they were determined from Eq.~(\ref{eq:coef_coupling}).

\begin{table}[ht!]
\centering
\caption{Fitting with polynomial dependence\label{tab:polynom}}
\begin{tabular}{c c c c c c c c c c c c}
\hline
$n$ & d & $w_{\mathrm{B}}$ & $w_{\mathrm{D}}$ & $c_2$ & $c_3$ & $c_4$ & $c_5$ & $c_6$ & $c_7$ & $c_8$ & $\mathcal{E}$ \\ \hline
2 & 95.8 & .656 & .293 & (1) & & & & & & & 1.27 \\
3 & 95.2 & .650 & .355 & (0) & 1.00 & & & & & & 1.23 \\
6 & 95.0 & .645 & .314 & (.754) & .418 & .491 & .617 & .770 & &  & 1.22 \\
8 & 96.1 & .638 & .381 & .681 & .369 & .399 & .504 & .354 & .706 & (.874) & 1.05 \\ \hline
\end{tabular}
\end{table}

\begin{figure}[ht!]
\centering\includegraphics[width=13cm]{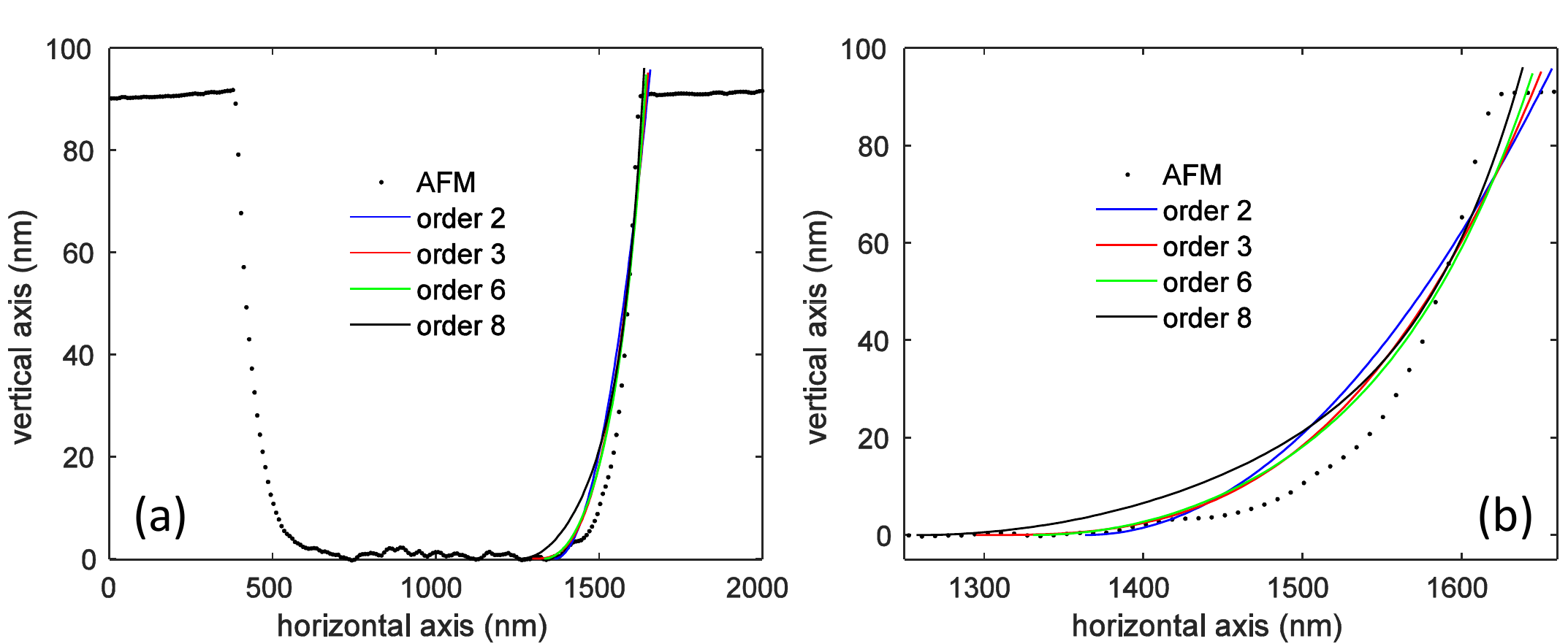}
\caption{Relief grating profile measured by AFM (dots) and patterned edge dependence numerically fitted by polynomials of 2nd, 3rd, 6th, and 8th orders (color curves). One period is displayed in (a), the detail of one edge in (b).}\label{fig:AFM_polynom}
\end{figure}

As clearly apparent from the table, increasing the order of the polynomial and thus the number of fitted variables reduces the error $\mathcal{E}$ of the SE fitted data from the value of $1.27^{\circ}$ down to the value of $1.05^{\circ}$. The bottom filling factor $w_{\mathrm{B}}$ also decreases toward the filling factor measured by AFM, as visible in Fig.~\ref{fig:AFM_polynom}.
Similarly, the fitted shape of the edge slightly approaches the edge detected by AFM, but a considerable difference remains even for high-order polynomials whose coefficients become for the least-square method mutually correlated.

\begin{figure}[ht!]
\centering\includegraphics[width=13cm]{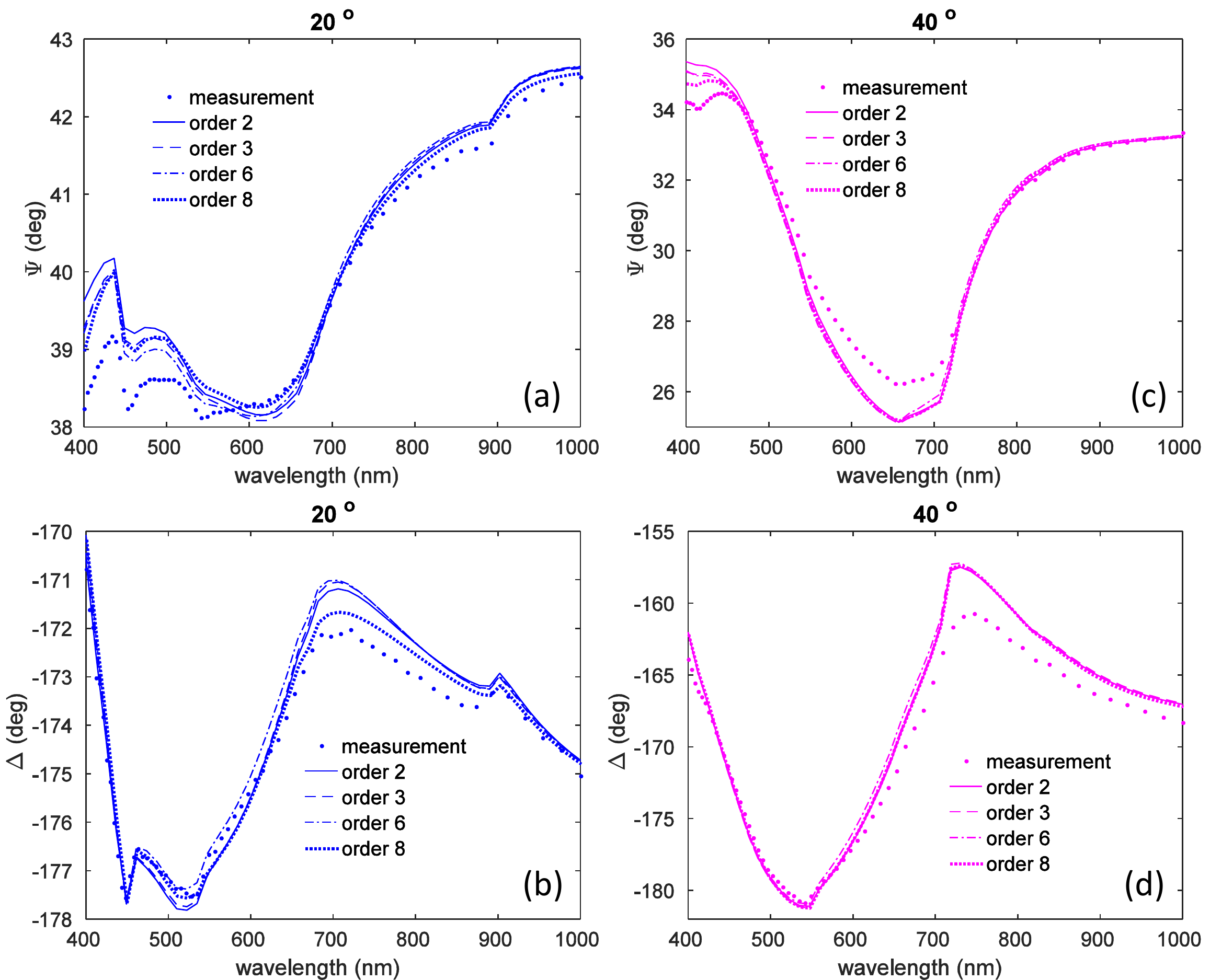}
\caption{Fitted ellipsometric parameters $\Psi$ (a,c) and $\Delta$ (b,d) for two chosen angles of incidence, $20^{\circ}$ (a,b) and $40^{\circ}$ (c,d) by polynomials of several orders. Experimental data are displayed by dots, fitted values by curves.}\label{fig:fit_polynom}
\end{figure}

The fitted ellipsometric parameters, $\Psi$ and $\Delta$, are displayed in Fig.~\ref{fig:fit_polynom} for two chosen angles of incidence, 20 and $40^{\circ}$. The tendency of the fitted curves suggests that no significant improvement can be expected if the order of the polynomial further increases.

To reduce the number of variable parameters but keep the shape of the edge, the polynomials were replaced by a power-dependent function with only one shape parameter,
\begin{equation}
    \frac{z}{d} = \left(\frac{y}{y_{\mathrm{D}}}\right)^{2+\alpha(y/y_{\mathrm{D}})},
\end{equation}
where $\alpha$ is the fitted ``power-increasing parameter.'' This function behaves as a quadratic function around the point $y=0$, and as a power function of order $2+\alpha$ around the point $y=y_{\mathrm{D}}$, so that in general it can replace a polynomial of the order $2+\alpha$. The first row of values in Table~\ref{tab:power} shows the result of the fitting of this function to the SE data.

\begin{table}[ht!]
\centering
\caption{Fitting with power function dependence\label{tab:power}}
\begin{tabular}{c c c c c c}
\hline
$n$, $k$ & $d$ & $w_{\mathrm{B}}$ & $w_{\mathrm{D}}$ & $\alpha$ & $\mathcal{E}$ \\ \hline
native & 96.2 & .642 & .385 & 2.06 & 1.08 \\
modified & 89.9 & .629 & .375 & 4.91 & 0.51 \\ \hline
\end{tabular}
\end{table}

\begin{figure}[ht!]
\centering\includegraphics[width=13cm]{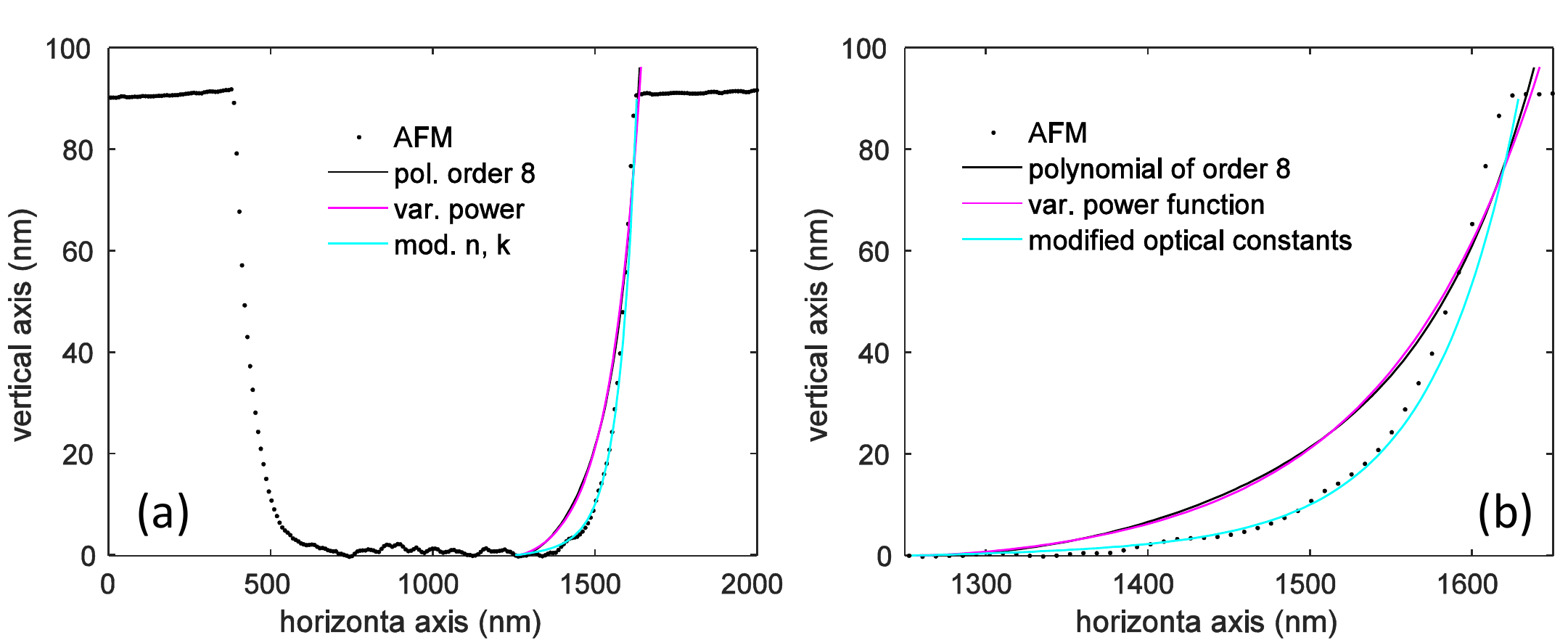}
\caption{Relief grating profile measured by AFM (dots) and patterned edge dependence numerically fitted by the polynomial of 8th order (black curve) and varied power function (magenta and cyan curves). The cyan curve corresponds to fitting with optical constants modified by EBL. One period is displayed in (a), the detail of one edge in (b).}\label{fig:AFM_final}
\end{figure}

The fitted geometric dimensions, the thickness $d$ and the filling factors $w_{\mathrm{B}}$ and $w_{\mathrm{D}}$ were found very close to those of the last row in Table~\ref{tab:polynom}, corresponding to the polynomial of 8th order, so that six fitted coefficients $c_j$ were replaced by only one, the power-increasing parameter~$\alpha$. The similarity of the both functions is clearly visible in Fig.~\ref{fig:AFM_final}, where the polynomial of 8th order is plotted by the black curve, whereas the power dependence function by the magenta curve.

So far all simulations were carried out with the native ChG optical constants measured on the unpatterned thin film reference sample without EBL modification. Since the studied ChG is treated as a negative resist, only the exposed part of the ChG layer remained on the sample after the EBL exposure and etching. Therefore, a new fit was carried out using modified optical constants that were previously measured on another $\mathrm{As_{35}S_{65}}$ sample modified by the same EBL procedure~\cite{Janicek_TSF_2018}. The modified optical constants are plotted in Fig.~\ref{fig:AsS_n_k} by the blue curves.

\begin{figure}[ht!]
\centering\includegraphics[width=13cm]{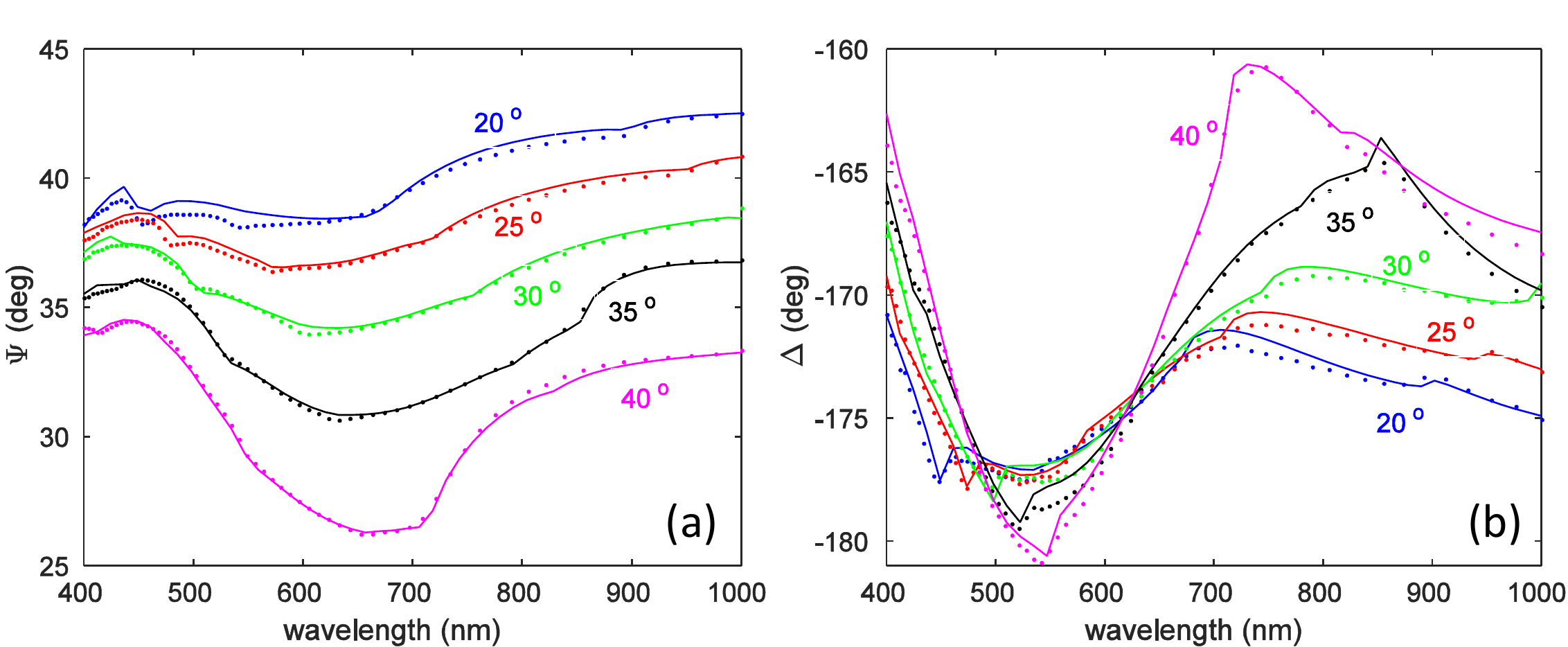}
\caption{Fitted ellipsometric parameters $\Psi$ (a) and $\Delta$ (b) for five measured angles of incidence ($20$, $25$, $30$, $35$, and $40^{\circ}$). Experimental data are displayed by dots, fitted values by curves.}\label{fig:fit_final}
\end{figure}

The result of the final fit is shown in the last row of Table~\ref{tab:power} and the fitted power-dependent function in Fig.~\ref{fig:AFM_final} (cyan color). Obviously, now the fitted grating profile is almost exactly same as the one detected by AFM, although the bottom linewidth is slightly larger. The ChG thickness $d=89.9$~nm almost precisely corresponds to the value measured by AFM. The error $\mathcal{E}=0.51^{\circ}$ is half of the one obtained when native ChG optical constants were used. The fitted ellipsometric spectra for all the five angles of incidence agree well with the experimental ones in the whole spectral range, as visible in Fig.~\ref{fig:fit_final}

Note that the nominally intended ChG film thickness ($d_0=100$~nm) slightly differs from the one detected experimentally when it was measured on the unpatterned reference film sample ($d_1=106$~nm), on the patterned grating by AFM ($d_2=90$~nm), and when it was obtained by fitting the SE experimental data (about $d_3=96$~nm assuming the native ChG optical constants and $d_4=89.9$~nm assuming the modified ones). The thickness of the reference sample was most probably detected accurately; its higher value could be caused by the deposition conditions which might not be precisely adjusted; the position of the sample in the deposition apparatus should also be taken into account. The value $d_3$ was obviously affected by the incorrect optical constants detected on the reference film sample which was not exposed to the EBL process. On the other hand, the value of $d_4$, for which the optical constants modified by EBL were assumed, is nearly the same as $d_2$ detected by AFM. We conclude that both $d_2\approx d_4$ are accurate.

The shape of the patterned lines, whose rounded edges were found approximately $y_{\mathrm{D}}=375$~nm wide, which is a size over four times higher than the ChG film thickness, can be explained by an imperfect focus of the electron beam used in EBL and by selectivity of the etching which was not precisely $100\ \%$. After recording the latent grating image on the ChG film, not only unexposed material of the film was removed but also the exposed area was partially etched out, yet considerably slower. This also accounts for the thickness of the patterned ChG lines, which is smaller than the thickness of the ChG film unexposed to EBL ($d_{2,4}<d_1$).

Finally, the quality of the grating sample was also investigated with respect to LER. All geometric and optical parameters were fixed at the final fitted values summarized in the last row of Table~\ref{tab:power}, and the parameter of quality, $\eta$, defined by Eqs.~(\ref{eq:LER1}) and~(\ref{eq:LER2}) was fitted. The result $\eta=1.00$ accounts for high quality of the pattern whose LER cannot be detected by SE in the spectral range used. The high quality pattern is also apparent from Fig.~\ref{fig:AFM_3D}.

\section{Conclusion}
A grating patterned on an $\mathrm{As_{35}S_{65}}$ ChG thin film was investigated by direct methods (optical microscopy, SEM, and AFM) and optical scatterometry based on SE. Geometric dimensions and the shape of the patterned lines were determined by fitting the SE data with high accuracy and considerably high agreement with the relief profile measured by AFM. The importance of correct ChG optical parameters used in the fitting procedure was demonstrated by decreasing the error of the final fit and improving the correspondence between the SE and AFM results. A very simple power function with linearly varied power with only one parameter of shape was successfully used to describe the edge dependence and thus the result of wet etching with high precision. The roughness of the line edges was investigated by a further fitting procedure with the result of the quality factor $\eta=1.00$, demonstrating high quality of the pattern whose LER cannot be detected by SE in the current configuration, which is a deciding factor for fabrication of optical tools. Therefore, SE is shown to be capable of determining the precise dimensions, shape, and roughness of patterned elements with high sensitivity, even if a relatively narrow spectral range of visible and near IR light is used.

\section*{Funding}
Grant Agency of the Czech Republic (GA~CR) (16-13876S).

%\section*{Acknowledgments}
%Acknowledgments, if included, should appear at the end of the document. The section title should not be numbered.

%%%%%%%%%%%%%%%%%%%%%%% References %%%%%%%%%%%%%%%%%%%%%%%%%

%%%%%%%%%% If using BibTeX:
\bibliography{Antos_biblio}

\end{document}